\documentclass{article}
\usepackage{spconf,amsmath,graphicx,hyperref}
\usepackage{xcolor}
\usepackage{tikz,pifont}
\usetikzlibrary{arrows.meta,positioning,calc}
\usepackage{booktabs}
\usepackage{multirow}
\usepackage{makecell}
\usepackage{array}
\usepackage{enumitem}
\usepackage{url,cite}
\usepackage{microtype}
\usepackage[moderate,lists=normal]{savetrees}

\mathchardef\UrlBreakPenalty=0
\mathchardef\UrlBigBreakPenalty=0

\usepackage[labelfont=bf,labelsep=period]{caption}
\makeatletter
\renewcommand\section{\@startsection{section}{1}{\z@}%
  {-2.0ex \@plus -1ex \@minus -.2ex}{1.2ex \@plus .2ex}%
  {\normalfont\Large\bfseries}}
\renewcommand\subsection{\@startsection{subsection}{2}{\z@}%
  {-2.0ex \@plus -1ex \@minus -.2ex}{0.8ex \@plus .2ex}%
  {\normalfont\large\bfseries}}
\makeatother
\DeclareMathOperator{\median}{median}
\newcommand{\Cstable}{\mathcal{S}}
\newcommand{\Ppool}{\mathcal{P}}
\newcommand{\Oqueue}{\mathcal{Q}}

\usepackage{algorithm,algorithmic}
\makeatletter\renewcommand{\ALG@name}{Algorithm}\makeatother

\newif\ifreview
\reviewfalse
\ifreview
  \usepackage{xetexko}
  \setmainhangulfont{Apple SD Gothic Neo}
  \definecolor{rvcbot}{RGB}{0,124,72}
  \definecolor{rvcbj}{RGB}{0,82,204}
  \definecolor{rvcmj}{RGB}{193,66,0}
  \newcommand{\rvnote}[4]{%
    \par\addvspace{2pt}%
    \noindent{\footnotesize\color{#1}%
      \textbf{[#2\,@#3]}~#4\par}\addvspace{2pt}}
  \newcommand{\rvBot}[3]{\rvnote{rvcbot}{C#1}{#2}{#3}}
  \newcommand{\rvBJ}[3]{\rvnote{rvcbj}{B#1}{#2}{#3}}
  \newcommand{\rvMJ}[3]{\rvnote{rvcmj}{M#1}{#2}{#3}}
  \definecolor{rvchs}{RGB}{122,30,168}
  \newcommand{\rvHS}[3]{\rvnote{rvchs}{H#1}{#2}{#3}}
  \newcommand{\rvfix}[1]{\par\nobreak
    {\normalcolor\footnotesize\textbf{\ding{52}\,반영(2026-09-08):}~#1\par}}
\else
  \newcommand{\rvBot}[3]{}
  \newcommand{\rvBJ}[3]{}
  \newcommand{\rvMJ}[3]{}
  \newcommand{\rvHS}[3]{}
  \newcommand{\rvfix}[1]{}
\fi

\title{TOWARD ACCURATE SPEAKER INVENTORIES\\FOR ONLINE SPEAKER DIARIZATION}

\name{Youngki Kwon \qquad Hee-Soo Heo \qquad Minjae Lee \qquad Bong-Jin Lee}
\address{NAVER Cloud Corporation, South Korea}

\begin{document}
\ninept
\maketitle

\begin{abstract}
An online speaker diarizer maintains a growing inventory of speakers, and
applications read it directly: a live transcript enumerates them as they
appear.  Existing systems either fix the inventory's size in advance, so
speakers beyond that size are merged, or grow it on distance alone, so noise
and turn boundaries split speakers.  Scored on speaker count over nine public
datasets, both failures appear where DER reports neither: on
VoxSRC\nobreakdash-23 the fixed-capacity systems return 44--46\% too few
speakers and the unbounded tracker with the lowest DER 63\% too many.  We
build on that embedding-based tracker and change only what triggers a
registration.  A candidate speaker pool keeps unregistered embeddings apart
and registers a speaker only once one candidate has gathered enough mutually
compatible embeddings; commit-gated label assignment holds an output while
its candidate is undecided, so no speaker appears in the output before it is
confirmed.  Across the nine datasets the macro speaker-count error falls from
225\% to 21\% of the reference and macro DER from 17.59\% to 15.95\%, at a
mean added latency of 0.109~s over all outputs; on VoxSRC\nobreakdash-23 the
tracker obtains both the lowest DER and the most accurate speaker count of
the systems compared.
\end{abstract}


\section{Introduction}
\label{sec:intro}

Speaker diarization determines ``who spoke when''
\cite{anguera2012speaker,park2022review}.  An online system must emit
temporally ordered labels under bounded latency.

Every online speaker diarizer maintains a growing set of speakers admitted so
far, which applications read directly; we call it the \emph{speaker
inventory}---built online, not enrolled in advance.  Prior work has no shared name
for it---it appears as attractors
\cite{horiguchi2022online,liang2025ls}, as a speaker cache
\cite{medennikov2025streaming}, as a speaker-tracing buffer
\cite{xue2021slt,xue2021online}, as centroids or retained samples
\cite{coria2021overlap,zhang2022low,yue2022online,kwon2023absolute}---and the
designs differ mainly in how they decide when to add to it.

Two kinds of rule govern how that set grows.  Some designs fix its size as an
architectural constant settled before the session---a \emph{speaker
capacity}, the most speakers the system can ever return: a number of output
slots \cite{medennikov2025streaming} or an attractor budget set by training
\cite{liang2025ls}; block-wise designs
\cite{horiguchi2022online,xue2021slt,xue2021online} apply that budget within
each block rather than to the whole recording.  Embedding-based
designs \cite{coria2021overlap,kwon2023absolute,zhang2022low,yue2022online}
leave it open and decide online: each incoming embedding is compared against the
inventory already held, and a speaker is admitted when nothing there is close
enough.  Either way the rule is fixed in advance, and once a label is
emitted it cannot be revised, so a registration taken on too little evidence
is never undone.

An inventory has a \emph{count error} when it returns a different number of
speakers than are present, in one of two directions.  Too few is a merge: two
or more speakers share one identity in the inventory, and the speech that
cannot be placed is charged to speaker confusion---what a fixed speaker
capacity produces once the true count passes it.  Too many is a split: one
speaker is spread over several identities---what unbounded registration
produces, since every sufficiently distant embedding becomes a speaker, and
conversation supplies them constantly.  We measure it simply, as the total absolute difference in
speaker count over the total reference count, both directions counted
(Section~\ref{sec:setup}).

DER, the primary metric of speaker diarization \cite{anguera2012speaker},
reports neither direction: it weights speakers by the time they hold and does
not state how many came back.  Among the existing systems on VoxSRC\nobreakdash-23
(Table~\ref{tab:baselines}), the lowest DER comes with the worst inventory, and
the most accurate inventory with more than twice that DER.  JER
\cite{ryant2019second}, another notable metric, weights every reference speaker
equally, which raises the weight of short speakers, but it is still scored over the optimal
speaker mapping and is silent on the count and its direction.

We follow the embedding-based designs, since only an inventory with no fixed
speaker capacity can hold the many speakers our target recordings contain.
We build on DC\nobreakdash-OST \cite{kwon2026dcost}, keeping its routing rule
and its delayed commitment (Section~\ref{sec:background}).  What we change
is what triggers a speaker registration (Section~\ref{sec:method}).  We compare against
existing systems on DIHARD~III and VoxSRC\nobreakdash-23, re-running the
released ones under the same protocol and scoring DER and speaker-count error
together.  The
proposed tracker is then analysed on nine datasets, component by component and
over its hyper-parameters, with the latency it costs
(Section~\ref{sec:results}).

Our contributions are:
\begin{itemize}[left=0pt,itemsep=1pt,topsep=2pt]
    \item A \emph{candidate speaker pool} that decides when the inventory
    grows (Section~\ref{sec:pool}).  A new speaker is registered only on
    accumulated evidence, kept separate per candidate.  The number of
    speakers registered therefore stays close to the number present: macro
    speaker-count error falls from 225\% to 21\% of the reference, and macro
    DER by 0.77\,\%p.
    \item \emph{Commit-gated label assignment} over it
    (Section~\ref{sec:delay}).  An output that may belong to a new speaker
    waits until the next few outputs confirm or reject that speaker.
    Look-ahead is therefore spent only where needed, not fixed on every
    output: macro DER falls a further 0.87\,\%p, at a mean added 0.109~s over
    all outputs---1.27~s over the 8.6\% held.
\end{itemize}

\section{Background}
\label{sec:background}

Our baseline is DC\nobreakdash-OST \cite{kwon2026dcost}, an embedding-based
tracker with no fixed speaker capacity.  Throughout,
$t$ indexes the embedding steps, taken every 0.5~s over detected speech only,
so consecutive steps can be far apart in time; $a_t$ is the audio time of step
$t$ and $\mathbf{e}_t$ the embedding of the window ending there.

The tracker holds a registry of committed centroids,
$\Cstable_t=\{\mathbf{c}_{t,i}\}_{i=1}^{N_t}$: $\mathbf{c}_{t,i}$ is the
centroid of speaker $i$ at step $t$, updated as the speaker accumulates
observations, and $N_t$ the number of registered speakers.  It also holds a
single buffer $B$ of embeddings that matched none of them.
It routes $\mathbf{e}_t$ by its nearest centroid under the cosine distance
$d$: with $d_{\min}=\min_{1\le i\le N_t} d(\mathbf{e}_t,\mathbf{c}_{t,i})$,
the embedding is assigned to that speaker if $d_{\min}\le\tau_t$ and appended
to $B$ otherwise, where $\tau_t$ grows linearly with $N_t$.

A label is emitted at once---the matched speaker's, or the next unused identifier
$N_t{+}1$ if $\mathbf{e}_t$ went to the buffer---and a speaker is registered
only when $|B|$ reaches a preset size $K$, with the medoid of $B$ as its
centroid.

\textbf{Limitations.}  Two properties of that procedure bound it.  (1)~One
buffer for every unseen speaker: two unseen speakers who alternate mix in
$B$, share one identifier, and are registered as one speaker with the medoid
of the mixture as its centroid.  (2)~Every decision from one observation: a
single embedding with $d_{\min}>\tau_t$ starts a new speaker, labelled
$N_t{+}1$ at once, so one outlier can seed a speaker who does not exist.

\section{Proposed Method}
\label{sec:method}

Section~\ref{sec:background} ends on two limitations; the proposed tracker
answers each.  The single buffer becomes a pool $\Ppool_t$ of candidate
speakers kept apart.  And no decision rests on one observation.  An embedding
is still placed at once, into a candidate, but the candidate becomes a
speaker only after enough mutually compatible embeddings have joined it.  An
output whose candidate is undecided is held in an ordered queue $\Oqueue_t$
until then: DC\nobreakdash-OST delays the registration, we delay the label as
well.  Nothing else in the pipeline changes.
Algorithm~\ref{alg:tracker} states one embedding step in full; the rest of
this section gives the reasons behind its two gates.

\begin{algorithm}[t]
\caption{One embedding step of the proposed tracker.  Lines 2--4 are the
common case, line 3 the routing test; the two gates are the tests on
$\tau_{\mathrm{pool}}$ (line 7) and $K$ (line 10).}
\label{alg:tracker}
\small
\begin{algorithmic}[1]
\item[\textbf{Input:}] embedding $\mathbf{e}_t$ at audio time $a_t$
\item[\textbf{Output:}] the labels released at step $t$: zero or more
        identifiers, each for the output held at audio time $a$, for
        $\mathbf{e}_t$ or for earlier steps still in $\Oqueue_t$
\item[\textbf{State:}] registry $\Cstable_t$ with $N_t=|\Cstable_t|$ speakers;
        pool $\Ppool_t$ of candidates $P_j$, each a set of embeddings
        $\mathbf{x}$ with their audio times, $\overline{P_j}$ their mean,
        $\mathrm{last}(P_j)$ the newest; pending
        queue $\Oqueue_t$ of entries $(a_t,\,i^{*},\,P^{*})$:
        audio time, fallback label, candidate ($\emptyset$ once decided)
\item[\textbf{Hyper-parameters:}] admission threshold $\tau_{\mathrm{pool}}$,
        registration size $K$, delay budget $L_{\mathrm{prov}}$,
        expiry $T_{\mathrm{e}}$, and $\tau_{\mathrm{base}}$, $\lambda$,
        $\tau_{\max}$ of the adaptive routing threshold
        $\tau_t=\min(\tau_{\mathrm{base}}+\lambda N_t,\,\tau_{\max})$
\STATE \textbf{if} $N_t=0$ \textbf{then} $N_t\leftarrow1$;
       $\mathbf{c}_{t,1}\leftarrow\mathbf{e}_t$
\STATE $i^{*}\leftarrow\arg\min_{1\le i\le N_t} d(\mathbf{e}_t,\mathbf{c}_{t,i})$;
       $d_{\min}\leftarrow d(\mathbf{e}_t,\mathbf{c}_{t,i^{*}})$
\IF{$d_{\min}\le\tau_t$}
  \STATE update $\mathbf{c}_{t,i^{*}}$ with $\mathbf{e}_t$; append
         $(a_t,\,i^{*},\,\emptyset)$ to $\Oqueue_t$
\ELSE
  \STATE $D_j\leftarrow \median(\{d(\mathbf{e}_t,\mathbf{x}):\mathbf{x}\in
         P_j\})$ for each $P_j\in\Ppool_t$
  \STATE \textbf{if} $\min_j D_j\le\tau_{\mathrm{pool}}$ \textbf{then}
         $P^{*}\leftarrow\arg\min_j D_j$; $P^{*}\leftarrow P^{*}\cup\{\mathbf{e}_t\}$
  \STATE \textbf{else} $P^{*}\leftarrow\{\mathbf{e}_t\}$;
         $\Ppool_t\leftarrow\Ppool_t\cup\{P^{*}\}$
  \STATE append $(a_t,\,i^{*},\,P^{*})$ to $\Oqueue_t$
  \IF{$|P^{*}|\ge K$}
    \STATE $N_t\leftarrow N_t+1$; $\mathrm{id}(P^{*})\leftarrow N_t$
    \STATE $\mathbf{c}_{t,N_t}\leftarrow\overline{P^{*}}/\|\overline{P^{*}}\|$;
           $\Ppool_t\leftarrow\Ppool_t\setminus P^{*}$
  \ENDIF
\ENDIF
\WHILE{$\Oqueue_t\neq\emptyset$ and its head $(a,\,i,\,P)$ has $P=\emptyset$,
       $P$ registered, or $a_t-a\ge L_{\mathrm{prov}}$}
  \STATE pop it; emit $\mathrm{id}(P)$ if $P$ registered, else $i$
\ENDWHILE
\STATE discard every $P_j\in\Ppool_t$ with $|P_j|=1$ and
       $a_t-\mathrm{last}(P_j)\ge T_{\mathrm{e}}$
\end{algorithmic}
\end{algorithm}

\subsection{Evidence-gated speaker registration}
\label{sec:pool}

The pool $\Ppool_t$ replaces $B$ with candidate speakers that stay apart, the
same $K$ now applying to each candidate separately rather than to one shared
buffer.  An embedding the registry rejects is scored against each candidate by
the median of its member-wise distances (line 6).  It joins the nearest
candidate if that distance is within the admission threshold
$\tau_{\mathrm{pool}}$, and opens a new one otherwise (lines 7--8).  Two unseen speakers who alternate therefore
accumulate in separate candidates instead of one buffer.

A candidate is registered once it holds $K$ members (line 10).  Because every
member joined by passing the admission test of line 7, the $K$ embeddings
are accumulated evidence for one speaker rather than $K$ embeddings that
happened to miss the registry.  The committed centroid is the
normalized mean of those members (line 12).  A candidate below
$K$ receives no label at all, so a single noisy observation leaves no trace in
the inventory, and beyond the first speaker (line 1) no speaker is registered
until one candidate holds $K$ of its embeddings; a candidate with $|P_j|=1$ after $T_{\mathrm{e}}$ of audio is
discarded (line 18).

\subsection{Commit-gated label assignment}
\label{sec:delay}

An output that joins an unresolved candidate is queued with both identities
it might take (line 9): a \emph{fallback}, the nearest speaker already in
$\Cstable_t$, and the \emph{provisional} identity of its candidate.  Whether
that candidate becomes a speaker is settled within a few steps, so the queue
holds the output instead of guessing.  The wait is timed out at
$L_{\mathrm{prov}}$, the delay budget, checked at each embedding step after
registrations are updated: the output leaves under the provisional identity
as soon as its candidate is registered, and under the fallback once its wait
reaches $L_{\mathrm{prov}}$ with the candidate still unregistered
(lines 15--16).  No speaker therefore
appears in the output before it is confirmed.  A timed-out output is not revisited; if its candidate registers
later, the error is a merge.
The queue releases in step order.  A routed output is decided on arrival
(line 4), so it waits only behind an earlier undecided one, never for
evidence of its own.  We call that wait \emph{head-of-line} blocking, as in
queueing: an output held back only by the one in front of it.
$L_{\mathrm{prov}}$ therefore
applies to the queue head, not to the queue.
\section{Experimental Setup}
\label{sec:setup}


\subsection{Datasets}

We evaluate on nine public datasets, 1{,}907 recordings.
\emph{DIHARD I--III} \cite{ryant2018first,ryant2019second,ryant2021third}
span eleven domains and are the hardest acoustic conditions of the nine: 1
to 9 speakers per recording, most of them under four, with moderate
overlap.
\emph{VoxConverse} \cite{chung2020spot} and \emph{VoxSRC 2021--2023}
\cite{huh2024vox} are in-the-wild video, and carry the many-speaker regime
this paper is about: a wide range of speakers, from 1 to 28, with the least
overlap of the nine.
\emph{AMI} \cite{carletta2005ami} and \emph{AliMeeting} \cite{yu2022m2met}
are English and Mandarin meetings---four speakers in all but one AMI
recording, two to four in AliMeeting---and carry the most overlap.
The comparison with existing systems (Table~\ref{tab:baselines}) uses the
evaluation sets of the two challenge series, VoxSRC\nobreakdash-23 and DIHARD III; our three variants
are compared on all nine.

\subsection{Evaluation protocol}

We report DER with its components---false alarm (FA), missed speech (MS) and
speaker confusion (SC)---in percent, with differences in percentage points
(\%p), and latency as a distribution (Section~\ref{sec:latency}).

\textbf{Overlap floor.}  Our tracker emits one label per step and misses
overlapped speech; the existing systems are scored on their own overlap-aware
output.  We therefore report each dataset's single-label missed-speech
floor, below which our MS cannot fall.

\textbf{Evaluation collar.}  Each dataset is scored under its own protocol---a
0.25-s collar on the Vox datasets, none elsewhere---so the per-dataset columns
stay comparable to published numbers.  A macro average over the nine would
then mix two protocols, so every macro average we report is computed after
rescoring all nine under one common 0.15-s collar.

\textbf{Latency.}  All latencies are algorithmic, not wall-clock.  An output
is one 0.5-s embedding step's label, and its \emph{added} latency is how long
the pending queue holds it
past the step at which the Baseline rule of Section~\ref{sec:systems} would
have emitted it.

\textbf{Speaker count error.}  For recording $i$, let $N^{\mathrm{sys}}_i$ be
the number of distinct speakers the system returns---a label counts if it is
emitted even once---and $N^{\mathrm{ref}}_i$ the number in the reference.
The signed error is $\varepsilon_i=N^{\mathrm{sys}}_i-N^{\mathrm{ref}}_i$: a merge
leaves a \emph{deficit} ($\varepsilon_i<0$) and a split an \emph{excess}
($\varepsilon_i>0$).  We report
$|e|=\sum_i|\varepsilon_i|/\sum_i N^{\mathrm{ref}}_i$ in percent, which
counts both alike instead of letting them cancel.  It measures the count,
not which identities are right; DER covers the time mislabelled.

\subsection{Implementation details}

Beyond two detector operating points, nothing below varies across the nine
datasets.

\textbf{Front-end module.}  The pipeline uses the public FireRedVAD
streaming detector \cite{xu2026fireredasr2s}---a 0.6M-parameter causal DFSMN
with zero look-ahead---and the released ReDimNet2 checkpoint
\texttt{b6-vb2+vox2\_v0-lm.pt} \cite{redimnet2_2026,artifact_redimnet2},
trained on VoxBlink2 and
VoxCeleb2, neither of them among the nine evaluation datasets; embeddings
from a 1.5-s window every 0.5~s give a base latency of 0.5~s.  The detector runs at one operating point per protocol family, chosen on the
DIHARD~III and VoxConverse development sets: trimmed segment ends where the
collar is zero, padded ones under the 0.25-s collar.  No evaluation dataset appears in the
detector's training description.

\textbf{Tracker.}  We use $K=5$, $\tau_{\mathrm{pool}}=0.45$,
$L_{\mathrm{prov}}=2.0$~s, singleton expiry $T_{\mathrm{e}}=30$~s (untuned), and, for the
adaptive routing threshold $\tau_t$ shared with the baseline
rule, $\tau_t=\min(\tau_{\mathrm{base}}+\lambda N_t,\,\tau_{\max})$ with
$(\tau_{\mathrm{base}},\lambda,\tau_{\max})=(0.45,0.005,0.8)$---all fixed on
the DIHARD~III and VoxConverse development sets as a trade-off between DER and
$|e|$; no evaluation set enters any of these choices.  The adaptive threshold is DC\nobreakdash-OST's mechanism
\cite{kwon2026dcost}, not one this paper introduces, so the sensitivity in
Section~\ref{sec:sensitivity} covers $\tau_{\mathrm{pool}}$ and $K$;
$L_{\mathrm{prov}}$ is swept in Section~\ref{sec:latency}.

\begin{table}[!t]
\centering
\caption{Comparison with existing systems (settings in
Section~\ref{sec:setup}).  Rows other than Baseline rule and
+\,Pool\,+\,Delay are our runs of the released artifacts, except
S2SND\nobreakdash-M$^\dagger$, which is quoted; published DIHARD~III DERs:
DIART 27.6, Streaming Sortformer 18.97, LS\nobreakdash-EEND 19.61.  Lat.\ = algorithmic latency (s); $\delta$ is the queue's added
delay: zero for 91.4\% of outputs, 1.27~s on average for the 8.6\% that wait
(0.109~s over all outputs), observed maximum 4.16~s
(Section~\ref{sec:latency}, Table~\ref{tab:delay}).  Bold = lowest DER and $|e|$ within each block.}
\label{tab:baselines}
\setlength{\tabcolsep}{2.0pt}
\begin{tabular}{lrrrrrr}
\toprule
System & FA & MS & SC & DER & $|e|$ & Lat. \\
\midrule
\multicolumn{7}{c}{\textit{VoxSRC\nobreakdash-23 ($\le$28 spk, MISS floor 1.6\%)}} \\
DIART \cite{coria2021overlap}             & 2.96 & \phantom{0}3.06 & 11.45 & 17.46 & 27.7 & 5.00 \\
Streaming Sortformer \cite{medennikov2025streaming} & 1.10 & \phantom{0}4.23 & 15.45 & 20.79 & 46.2 & 1.04 \\
LS\nobreakdash-EEND \cite{liang2025ls}                & 2.62 & \phantom{0}8.40 & 21.49 & 32.51 & 43.7 & 1.00 \\
Baseline rule                             & 1.87 & \phantom{0}2.48 & \phantom{0}3.05 & \phantom{0}7.40 & 63.1 & 0.50 \\
+\,Pool\,+\,Delay                         & 1.87 & \phantom{0}2.48 & \phantom{0}2.41 & \phantom{0}\textbf{6.76} & \textbf{16.0} & $0.50{+}\delta$ \\
\midrule
\multicolumn{7}{c}{\textit{DIHARD III (MISS floor 9.5\%)}} \\
DIART \cite{coria2021overlap}             & 5.83 & \phantom{0}9.65 & 13.06 & 28.53 & \phantom{0}27.0 & 5.00 \\
Streaming Sortformer \cite{medennikov2025streaming} & 6.15 & \phantom{0}9.68 & \phantom{0}3.93 & 19.76 & \phantom{0}18.7 & 1.04 \\
LS\nobreakdash-EEND \cite{liang2025ls}                & 6.13 & \phantom{0}9.77 & \phantom{0}5.49 & 21.39 & \phantom{0}\textbf{14.5} & 1.00 \\
S2SND\nobreakdash-M$^\dagger$ \cite{cheng2025sequence} & -- & -- & -- & \textbf{17.12} & -- & 0.80 \\
Baseline rule                             & 3.98 & 13.41 & 10.89 & 28.28 & 294.6 & 0.50 \\
+\,Pool\,+\,Delay                         & 3.98 & 13.41 & \phantom{0}6.92 & 24.31 & \phantom{0}23.2 & $0.50{+}\delta$ \\
\bottomrule
\end{tabular}
\end{table}

\subsection{Systems compared}
\label{sec:systems}

We compare three variants of our own, three released systems run at their
published settings, and one we quote.

\textbf{Our variants.}  The three share the front end and segmentation and
differ in decision policy: \textbf{Baseline rule}, i.e.\ the routing rule of
Section~\ref{sec:background} with one shared buffer at the same $K$
\cite{kwon2026dcost}; \textbf{+\,Pool} (Section~\ref{sec:pool}); and
\textbf{+\,Delay} (Section~\ref{sec:delay}).  The Baseline rule registers the
medoid of $B$ as the centroid, the other two the normalized mean of a
candidate's members.
Every variant labels every detected output---+\,Pool releases an undecided
output at once under its fallback ($L_{\mathrm{prov}}=0$), and the queue is
flushed at the end of the stream---so FA and MS are identical across
variants and every difference is speaker confusion.

\textbf{Existing systems.}  They govern the inventory differently.
\textbf{Streaming Sortformer}
\cite{medennikov2025streaming} has four output slots backed by an
arrival-order cache, so a fifth speaker overwrites the oldest instead of
receiving an identity; we run the released 4-speaker
checkpoint \cite{artifact_sortformer} at its published 1.04-s setting.
\textbf{LS\nobreakdash-EEND} \cite{liang2025ls} fixes the attractor sequence to a
constant length $S\,{=}\,8$ and reads the active count below it from a
termination marker; we use the official DIHARD~III checkpoint
\cite{artifact_lseend} at 1~s, no Vox-tuned checkpoint being released.
\textbf{DIART} \cite{coria2021overlap} imposes no speaker capacity, holding one
representation per speaker and creating a speaker when an embedding is
farther than a constant from all of them; we run it at its default 5-s
latency with the published hyper-parameters \cite{artifact_diart}, the
VoxConverse setting on VoxSRC\nobreakdash-23 and the DIHARD~III setting on DIHARD~III.
\textbf{S2SND\nobreakdash-M} \cite{cheng2025sequence} releases no artifact, so we quote
its published DIHARD~III number.

\section{Results}
\label{sec:results}
\begin{figure}[t]
\centering
\includegraphics[width=\linewidth]{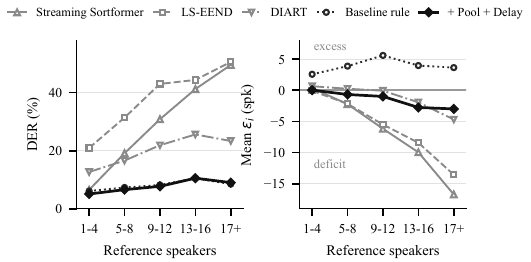}
\caption{Per-speaker-count behaviour on VoxSRC\nobreakdash-23, grouped by the number of
reference speakers (197/128/58/17/13 recordings).  Left: DER.  Right: mean
signed count error $\varepsilon_i$ (Section~\ref{sec:setup}), speakers
returned minus speakers present; zero means no net count bias.}
\label{fig:buckets}
\end{figure}
\begin{table*}[t]
\centering
\caption{Per-dataset results, both axes and both segmentation conditions.
SC and DER as baseline rule\,/\,pool\,/\,pool with delay, bold = best;
$|e|$ (\%) as baseline rule\,/\,pool, the pool and the full tracker
returning identical inventories.  FA and MS are shared by the three
variants; under reference boundaries FA is 0.00 throughout.  Macro: all nine
rescored under one 0.15-s collar (Section~\ref{sec:setup}), so it is not the
mean of the DER columns.}
\label{tab:main}
\setlength{\tabcolsep}{2.0pt}
\begin{tabular}{@{}c@{\hspace{3pt}}lrrccc rccc@{}}
\toprule
& & \multicolumn{5}{c}{\textit{detected speech}}
& \multicolumn{4}{c}{\textit{reference boundaries}} \\
\cmidrule(lr){3-7}\cmidrule(lr){8-11}
\textit{collar} & Dataset & \multicolumn{1}{c}{FA} & \multicolumn{1}{c}{MS}
& SC & DER & $|e|$ & \multicolumn{1}{c}{MS} & SC & DER & $|e|$ \\
\midrule
\multirow{5}{*}{\textit{none}}
 & DIHARD I & 5.31 & 13.58 & 12.58/9.16/\textbf{8.55} & 31.47/28.05/\textbf{27.45} & 278.4/21.4 & \phantom{0}8.71 & 15.17/10.30/\textbf{9.60} & 23.88/19.01/\textbf{18.32} & 329.2/24.4 \\
 & DIHARD II & 5.83 & 14.54 & 13.49/9.56/\textbf{8.99} & 33.86/29.92/\textbf{29.36} & 281.6/23.8 & \phantom{0}9.69 & 15.76/10.36/\textbf{9.71} & 25.45/20.05/\textbf{19.40} & 323.9/29.1 \\
 & DIHARD III & 3.98 & 13.41 & 10.89/7.34/\textbf{6.92} & 28.28/24.73/\textbf{24.31} & 294.6/23.2 & \phantom{0}9.52 & 12.31/7.42/\textbf{6.92} & 21.83/16.94/\textbf{16.44} & 331.9/25.2 \\
 & AMI & 1.95 & 26.32 & 6.13/3.66/\textbf{3.46} & 34.40/31.93/\textbf{31.74} & 466.7/47.6 & 14.55 & 7.44/4.59/\textbf{4.35} & 21.99/19.14/\textbf{18.90} & 460.3/50.8 \\
 & AliMeet. & 0.56 & 28.30 & \textbf{5.74}/6.00/5.90 & \textbf{34.60}/34.86/34.76 & 440.0/13.3 & 20.44 & 5.87/4.51/\textbf{4.40} & 26.31/24.95/\textbf{24.84} & 413.3/16.7 \\
\cmidrule(lr){2-11}
\multirow{4}{*}{\textit{0.25 s}}
 & VoxConv. & 1.91 & \phantom{0}2.75 & 3.05/3.23/\textbf{2.06} & 7.71/7.90/\textbf{6.73} & 82.7/14.0 & \phantom{0}1.60 & 3.10/3.23/\textbf{1.86} & 4.70/4.83/\textbf{3.46} & 67.7/14.4 \\
 & VoxSRC\nobreakdash-21 & 1.45 & \phantom{0}2.30 & 3.08/3.95/\textbf{2.42} & 6.83/7.70/\textbf{6.16} & 59.5/12.1 & \phantom{0}1.69 & 3.22/3.95/\textbf{2.21} & 4.91/5.63/\textbf{3.89} & 48.4/11.4 \\
 & VoxSRC\nobreakdash-22 & 1.83 & \phantom{0}2.23 & 2.80/3.67/\textbf{2.18} & 6.86/7.72/\textbf{6.24} & 58.0/14.7 & \phantom{0}1.54 & 2.92/3.73/\textbf{2.04} & 4.46/5.27/\textbf{3.58} & 48.2/13.6 \\
 & VoxSRC\nobreakdash-23 & 1.87 & \phantom{0}2.48 & 3.05/3.78/\textbf{2.41} & 7.40/8.14/\textbf{6.76} & 63.1/16.0 & \phantom{0}1.64 & 3.22/3.83/\textbf{2.28} & 4.86/5.47/\textbf{3.92} & 54.3/16.0 \\
\midrule
\textit{0.15 s} & Macro & 1.99 & \phantom{0}9.66 & 5.95/5.17/\textbf{4.31} & 17.59/16.82/\textbf{15.95} & 225.0/20.7 & \phantom{0}6.08 & 6.62/5.26/\textbf{4.27} & 12.71/11.35/\textbf{10.35} & 230.8/22.4 \\
\bottomrule
\end{tabular}
\end{table*}


\subsection{Comparison with existing systems}
\label{sec:external}

On VoxSRC\nobreakdash-23, +\,Pool\,+\,Delay has both the lowest DER and the
lowest $|e|$ of the systems compared (Table~\ref{tab:baselines}); the
Baseline rule is within 0.64\,\%p of its DER but returns 63.1\% too many
speakers.  On DIHARD~III our system trails Streaming Sortformer by 4.6\,\%p of
DER---3.0 of confusion, 1.6 of FA${}+{}$MS---and LS\nobreakdash-EEND on
$|e|$.

Fig.~\ref{fig:buckets} splits VoxSRC\nobreakdash-23 by reference speaker
count.  By DER the two embedding-based trackers, Baseline rule and
+\,Pool\,+\,Delay, are indistinguishable in every group, while their
inventories diverge from the first group on---the axis DER does not report.
Streaming
Sortformer returns at most four speakers, so beyond four its deficit is
$N^{\mathrm{ref}}-4$ by construction, and its DER reaches 50\% in the 17+
group.  DIART carries no capacity and still ends short there.
+\,Pool\,+\,Delay is near-exact up to four speakers and ends 3.0 short in the
17+ group, against 13.5 to 16.7 for the fixed-capacity systems.

None of the existing systems was trained on the VoxSRC\nobreakdash-23 test
data; DIART and LS\nobreakdash-EEND run checkpoints from other domains there,
while Streaming Sortformer's training includes VoxConverse dev.  On
DIHARD~III a fixed capacity helps when a recording has fewer speakers than
it: both
fixed-capacity systems count DIHARD~III more accurately than our system does.

\subsection{Both axes across nine datasets}
\label{sec:main}
\label{sec:mechanisms}

Table~\ref{tab:main} presents the paper's main result: on nine datasets both
axes improve---the macro count error by an order of magnitude, DER on eight
of the nine.  The
pool does most
of the work wherever unseen speakers alternate in one buffer---confusion
falls 3.55\,\%p on DIHARD~III and 2.47\,\%p on AMI---and the delay adds little
there.  On the Vox datasets the ordering reverses: the pool alone
\emph{raises} confusion, by 0.18 to 0.87\,\%p, and the delay removes more
than it added, 1.17 to 1.53\,\%p, because without the delay a genuinely new
speaker's first segments leave under a fallback label.  AliMeeting is the one dataset where the pool does not help, and it carries
the highest overlap floor: sustained two-speaker overlap yields recurring,
mutually compatible \emph{mixture} embeddings, the one input the evidence
gate cannot reject.  DER there is flat, 34.60 to 34.76, while $|e|$ goes
440\% to 13\%.

On the no-collar datasets 69--89\% of the DER is FA and MS.  To see the tracker without the detector's errors we
rescore with reference boundaries: there DER improves on all nine datasets,
and the residual MS is exactly each dataset's single-label floor.  The count errors barely
move: macro 225.0 to 230.8 and 20.7 to 22.4, the rises coming from the three
DIHARD sets.

\begin{table}[t]
\centering
\caption{Sensitivity to the two tracker hyper-parameters, macro DER and $|e|$ (\%)
under the common 0.15-s collar; $\star$ = the setting we run, bold = best in
row.  For $K$, DER is given with the budget fixed at $L_{\mathrm{prov}}=2.0$~s
and matched to $K$, $(K{-}1)\times0.5$~s; latency is the mean added latency
of the matched run.}
\label{tab:ablation}
\setlength{\tabcolsep}{2.4pt}
\begin{tabular}{@{}lrrrrrrr@{}}
\toprule
$\tau_{\mathrm{pool}}$ ($K{=}5$) & 0.30 & 0.35 & 0.40 & \llap{$\star$\,}0.45 & 0.50 & 0.55 & 0.60 \\
\midrule
DER   & 19.03 & 17.44 & 16.24 & 15.95 & \textbf{15.77} & 16.06 & 16.56 \\
$|e|$ & 25.1 & 21.5 & \textbf{20.3} & 20.7 & 25.9 & 37.2 & 55.6 \\
\midrule
$K$ ($\tau_{\mathrm{pool}}{=}0.45$) & 3 & 4 & \llap{$\star$\,}5 & 6 & 7 & & \\
\midrule
DER, fixed   & 16.18 & \textbf{15.82} & 15.95 & 16.59 & 16.97 & & \\
DER, matched & 16.18 & \textbf{15.82} & 15.95 & 16.44 & 16.70 & & \\
$|e|$                                 & 63.6 & 29.3 & 20.7 & \textbf{19.5} & 20.5 & & \\
latency (s)  & 0.032 & 0.068 & 0.109 & 0.154 & 0.199 & & \\
\bottomrule
\end{tabular}
\end{table}

\subsection{Sensitivity}
\label{sec:sensitivity}

Table~\ref{tab:ablation} varies each hyper-parameter around the operating
point of Section~\ref{sec:setup}, one at a time over the nine-dataset macro
average.  Neither is best on both axes (bold), and what we run sits between
them---0.18 and 0.13\,\%p off each sweep's DER optimum, buying 5.2 and
8.6\,\%p of $|e|$ for it.  Across both sweeps DER moves within 3.3\,\%p while $|e|$ swings
from 19.5 to 63.6---which DER does not report---yet stays far below the
baseline rule's 225.0.  The $K$ block also gives DER with $L_{\mathrm{prov}}$
set to $(K{-}1)\times0.5$~s instead of 2.0~s: $K{=}3$ and 4 are unchanged,
and $K{=}6$ and 7 recover 0.15 and 0.27\,\%p and still trail $K{=}5$.

\begin{table}[t]
\centering
\caption{The delay budget, over all 1,473,619 outputs of the nine datasets.
(a) $L_{\mathrm{prov}}$
sweep: macro SC (common 0.15-s collar) against the mean added latency over
all outputs; $\star$ = the setting we run.  (b) The 8.6\% the queue holds,
by exit (Section~\ref{sec:delay}); \% is of all outputs, mean and max added
latency (s) are over the row.}
\label{tab:delay}
\setlength{\tabcolsep}{2.6pt}
\begin{tabular}[t]{@{}c@{\hspace{6pt}}rr@{}}
\toprule
\multicolumn{3}{c}{\textit{(a) $L_{\mathrm{prov}}$ sweep}} \\
$L_{\mathrm{prov}}$ (s) & mean (s) & SC (\%) \\
\midrule
0.0 & 0.000 & 5.17 \\
0.5 & 0.027 & 4.91 \\
1.0 & 0.054 & 4.69 \\
1.5 & 0.081 & 4.49 \\
\llap{$\star$\,}2.0 & 0.109 & \textbf{4.31} \\
2.5 & 0.136 & \textbf{4.31} \\
\bottomrule
\end{tabular}%
\hspace{1.6em}%
\begin{tabular}[t]{@{}lrrr@{}}
\toprule
\multicolumn{4}{c}{\textit{(b) the 8.6\% that wait}} \\
exit & \% & mean & max \\
\midrule
provisional  & 2.22 & 0.76 & 2.00 \\
fallback     & 2.44 & 2.02 & 4.16 \\
head-of-line & 3.90 & 1.09 & 3.42 \\
\midrule
all          & 8.56 & 1.27 & 4.16 \\
\bottomrule
\end{tabular}
\end{table}

\subsection{The delay budget and what it costs}
\label{sec:latency}

To place $L_{\mathrm{prov}}$ we sweep it, measuring macro SC against the
mean added latency actually incurred (Table~\ref{tab:delay}a); Table~\ref{tab:delay}b
splits the outputs the queue holds by how they leave.  The sweep saturates at
$(K{-}1)$ embedding steps---the time a candidate needs to gather its $K$
members---exactly the $L_{\mathrm{prov}}=2.0$~s we adopt: a further 0.5~s
buys nothing.

The front-end bounds are shared by every variant, so the tracker's entire cost
is the hold.  It touches 8.6\% of outputs, a mean added 0.109~s over all
outputs and 1.27~s over those held (Table~\ref{tab:delay}b): 2.2\% resolve to
their own candidate, 2.4\% time out
and leave under the fallback---the only guesses---and 3.9\% were already
decided and waited only behind an earlier undecided output, the head-of-line
cost of releasing in step order.  The 4.16-s maximum is a timeout that fell
in a pause in speech, released at the next embedding step.

\section{Conclusion}

No system we measured keeps the speaker inventory accurate across speaker
counts, and DER does not report it, so we report the speaker-count error
alongside it.  We kept the count open and let no decision rest on one
observation: a candidate speaker pool registers a speaker only on
accumulated evidence, and commit-gated label assignment holds an output
while its candidate is undecided.  Across nine datasets this
cuts the macro speaker-count error from 225\% to 21\% and macro DER by
1.64\,\%p, at a mean added 0.109~s over all outputs;
the count is settled by the pool alone, the delay only reallocates labels.

The clearest limitation is overlap: one label per step misses overlapped
speech, four fifths of the meeting datasets' residual after
segmentation.  Admitting more than one speaker per step is next.\label{ENDBODY}

\bibliographystyle{IEEEbib}
\bibliography{strings,refs}

\end{document}